%% file: placco.tex
\documentclass[twocolumn]{aastex62} 

\usepackage{hyperref,enumerate}
\input{colordvi}

\hypersetup{
    unicode=false,                  
    pdftoolbar=true,                
    pdfmenubar=true,                
    pdffitwindow=true,              
    pdfstartview={FitH},            
    pdftitle={R443130},             
    pdfauthor={vplacco},            
    pdfsubject={Astronomy},         
    pdfcreator={dvipdf},            
    pdfproducer={dvipdf},           
    pdfkeywords={metal-poor stars}, 
    pdfnewwindow=true,              
    colorlinks=true,                
    linkcolor=red,                  
    citecolor=blue,                 
    filecolor=magenta,              
    urlcolor=cyan,                  
    breaklinks=true,
    linktocpage
}

\newcommand{\xx}{{\tablenotemark{a}}}
\newcommand{\yy}{{\tablenotemark{b}}}
\newcommand{\zz}{{\tablenotemark{c}}}
\newcommand{\abund}[2]{\ensuremath{[\mathrm{#1}/\mathrm{#2}]}}
\newcommand{\afe}{\abund{\alpha}{Fe}}
\newcommand{\cfe}{\abund{C}{Fe}}

\newcommand{\xfe}[1]{\abund{#1}{Fe}}
\newcommand{\metal}{\abund{Fe}{H}}

\newcommand{\teff}{\ensuremath{T_\mathrm{eff}}}
\newcommand{\logg}{\ensuremath{\log\,g}}

\newcommand{\pc}{\ensuremath{\%}}

\shorttitle{Metal-Poor Stars from the B\&B Survey}
\shortauthors{Placco et al.}

\begin{document}

\title{The \emph{R}-Process Alliance: Spectroscopic Follow-up of
Low-Metallicity \\
Star Candidates from the Best \& Brightest Survey}

\author[0000-0003-4479-1265]{Vinicius M.\ Placco}
\altaffiliation{Visiting astronomer, Kitt Peak National Observatory}
\affiliation{Department of Physics, University of Notre Dame, Notre Dame, IN 46556, USA}
\affiliation{JINA Center for the Evolution of the Elements, USA}

\author[0000-0002-7529-1442]{Rafael M.\ Santucci}
\affiliation{Instituto de Estudos S\'ocio-Ambientais, Planet\'ario, 
Universidade Federal de Goi\'as, Goi\^ania, GO 74055-140, Brazil}
\affiliation{Instituto de F\'isica, Universidade Federal de Goi\'as, Campus
Samambaia, Goi\^ania, GO 74001-970, Brazil}

\author[0000-0003-4573-6233]{Timothy C.\ Beers}
\affiliation{Department of Physics, University of Notre Dame, Notre Dame, IN 46556, USA}
\affiliation{JINA Center for the Evolution of the Elements, USA}

%

\author{Julio Chanam\'e}
\affiliation{Instituto de Astrof\'isica, Pontificia Universidad Cat\'olica de
Chile, Santiago, Chile}
\affiliation{Millenium Institute of Astrophysics, Santiago, Chile}

\author{Mar\'ia Paz Sep\'ulveda}
\affiliation{Instituto de Astrof\'isica, Pontificia Universidad Cat\'olica de
Chile, Santiago, Chile}
\affiliation{Millenium Institute of Astrophysics, Santiago, Chile}

\author{Johanna Coronado}
\affiliation{Max-Planck-Institut f\"ur Astronomie, 
K\"oningstuhl 17, D-69117 Heidelberg, Germany}
\affiliation{Instituto de Astrof\'isica, Pontificia Universidad Cat\'olica de
Chile, Santiago, Chile}

\author[0000-0001-7479-5756]{Silvia Rossi}
\affiliation{Instituto de Astronomia,  Geof\'{i}sica e Ci\^{e}ncias Atmosf\'{e}ricas,
Universidade de S\~{a}o Paulo, SP 05508-900, Brazil}

\author{Young Sun Lee}
\affiliation{Department of Astronomy and Space Science, Chungnam National
University, Daejeon 34134, Korea}

\author{Else Starkenburg}
\affiliation{Leibniz Institute for Astrophyics Potsdam, 
D-14482 Potsdam, Germany}

\author{Kris Youakim}
\affiliation{Leibniz Institute for Astrophyics Potsdam, 
D-14482 Potsdam, Germany}

\author{Manuel Barrientos}
\affiliation{Instituto de Astrof\'isica, Pontificia Universidad Cat\'olica de
Chile, Santiago, Chile}
\affiliation{Millenium Institute of Astrophysics, Santiago, Chile}

\author{Rana Ezzeddine}
\affiliation{JINA Center for the Evolution of the Elements, USA}
\affiliation{Department of Physics and Kavli Institute for Astrophysics and
Space Research, \\ Massachusetts Institute of Technology, Cambridge, MA 02139, USA}

\author[0000-0002-2139-7145]{Anna Frebel}
\affiliation{Department of Physics and Kavli Institute for Astrophysics and
Space Research, \\ Massachusetts Institute of Technology, Cambridge, MA 02139, USA}
\affiliation{JINA Center for the Evolution of the Elements, USA}

\author[0000-0001-6154-8983]{Terese T.\ Hansen}
\affiliation{Observatories of the Carnegie Institution for Science, 
Pasadena, CA 91101, USA}

\author[0000-0002-5463-6800]{Erika M.\ Holmbeck}
\affiliation{Department of Physics, University of Notre Dame, Notre Dame, IN 46556, USA}
\affiliation{JINA Center for the Evolution of the Elements, USA}

\author[0000-0001-6154-8983]{Alexander P.\ Ji}
\altaffiliation{Hubble Fellow}
\affiliation{Observatories of the Carnegie Institution for Science, 
Pasadena, CA 91101, USA}

\author[0000-0002-0470-0800]{Kaitlin C.\ Rasmussen}
\affiliation{Department of Physics, University of Notre Dame, Notre Dame, IN 46556, USA}
\affiliation{JINA Center for the Evolution of the Elements, USA}

\author[0000-0001-5107-8930]{Ian U.\ Roederer}
\affiliation{Department of Astronomy, University of Michigan, Ann Arbor, MI 48109, USA}
\affiliation{JINA Center for the Evolution of the Elements, USA}

\author[0000-0002-5095-4000]{Charli M.\ Sakari}
\affiliation{Department of Astronomy, University of Washington, Seattle, WA 98195-1580, USA}

\author[0000-0002-9594-6143]{Devin D.\ Whitten}
\affiliation{Department of Physics, University of Notre Dame, Notre Dame, IN 46556, USA}
\affiliation{JINA Center for the Evolution of the Elements, USA}

\correspondingauthor{Vinicius M.\ Placco}
\email{vplacco@nd.edu}

\begin{abstract}

We present results from an observing campaign to identify
low-metallicity stars in the Best \& Brightest Survey. From medium-resolution
($R \sim 1,200 - 2,000$) spectroscopy of 857 candidates, we estimate the
stellar atmospheric parameters (\teff, \logg, and \metal), as well as carbon and
$\alpha$-element abundances.
We find that 69\% of the observed stars have \metal\,$\leq -1.0$, 39\% have
\metal\,$\leq -2.0$, and 2\% have \metal\,$\leq -3.0$. 
There are also 133 carbon-enhanced metal-poor (CEMP) stars in this sample, with
97 CEMP Group~I and 36 CEMP Group~II stars identified in the $A$(C) versus [Fe/H] diagram.  
A subset of the confirmed low-metallicity stars were
followed-up with high-resolution spectroscopy, as part of the {\emph R}-process
Alliance, with the goal of identifying new highly and moderately $r$-process-enhanced stars.
Comparison between the stellar atmospheric parameters estimated in
this work and from high-resolution spectroscopy exhibit good agreement, confirming our
expectation that medium-resolution observing campaigns are an effective way
of selecting interesting stars for further, more targeted, efforts.

\vspace{2.50cm}

\end{abstract}


\section{Introduction}
\label{intro}

Very metal-poor (VMP; [Fe/H]\footnote{\abund{A}{B} = $\log(N_A/{}N_B)_{\star} -
\log(N_A/{}N_B) _{\odot}$, where $N$ is the number density of atoms of a given
element in the star ($\star$) and the Sun ($\odot$), respectively.}\,$< -2.0$)
stars are the {\emph{Rosetta Stones}} of stellar astrophysics in the early
Universe. Encoded in the atmosphere of these low-mass, long-lived stars are the
signatures of nucleosynthetic processes that could have occurred as early as a
few tens of million years after the Big Bang \citep{alvarez2006}.  This provides
a unique opportunity to witness not only the chemical and dynamical evolution of
the Milky Way, but also to identify and distinguish between a number of possible
scenarios for the enrichment of early star-forming gas clouds
\citep{jeon2017,chiaki2018}.

It has long been recognized that metal-poor stars with over-abundances of carbon
relative to iron (\cfe\ $ > +0.7$) become more frequent for decreasing
metallicities. The fractions of carbon-enhanced metal-poor (CEMP) stars increase
from 15-20\% for VMP stars to more than 80\% for ultra metal-poor (UMP; \metal$<
-4.0$) stars \citep{lee2013,yong2013b,placco2014c,yoon2018}.  The elemental abundance
patterns of CEMP stars are required in order to probe the nature of different
progenitor populations responsible for the production of carbon and other
elements \citep{placco2015,placco2016b}. Moreover, recent studies
\citep{norris2013b} show that the majority of CEMP stars with \metal\ $ < -3.0$
belong to the CEMP-no sub-class, characterized by the lack of enhancements in
their neutron-capture elements (\xfe{Ba} $ < 0.0$). The brightest extremely
metal-poor (EMP; \metal$< -3.0$) star in the sky, BD$+$44$^\circ$493, with
\metal=$-3.8$ and $V=9.1$, is a CEMP-no star \citep{ito2013}, and shares a
common elemental-abundance signature with the recently discovered CEMP-no star
with \metal\ $ \lesssim -8.0$ \citep{keller2014,bessell2015,nordlander2017}.
This distinctive CEMP-no pattern has also been identified in high-$z$ damped
Lyman-$\alpha$ systems \citep{cooke2012}, and is common among stars in
ultra-faint dwarf galaxies, such as SEGUE-1 \citep{frebel2014}. These
observations suggest that CEMP-no stars exhibit the nucleosynthesis products of
the very first generation of stars \citep{hansen2016,chiaki2017,hartwig2018}.

Bright VMP stars {\it without} C-enhancement are ideal targets for the
high-resolution spectroscopic identification of new examples of the rare class
of stars with moderate-to-large enhancements of elements associated with the
rapid neutron-capture process ($r$-process), the so-called $r$-I ($+0.3 \leq $
\abund{Eu}{Fe} $ \leq +1.0$ and \abund{Ba}{Eu}$ < 0.0$) and $r$-II
(\abund{Eu}{Fe} $ > +1.0$ and \abund{Ba}{Eu}$ < 0.0$) stars, respectively
\citep{beers2005,frebel2018}.  Until recently, only $\sim 25$ $r$-II stars had
been identified after their recognition some 25 years ago
\citep{sneden1994,sneden1996}.  Characterization of additional examples of such
stars is crucial in order to explore the origin of the astrophysical
$r$-process, constrain the nature of their likely progenitor(s) \citep[e.g., neutron
star mergers,][]{abbott2017,drout2017,shappee2017}, and to measure the
abundances of the radioactive chronometers thorium and uranium, which are only
presently available for a handful of stars.

The {\emph{R}}-Process Alliance (RPA) was recently established to fulfill the
need for further observational constraints on the astrophysical origin of the
$r$-process. 
Its overall science goal is to support stellar archaeology and nuclear
astrophysics by identifying as many $r$-process-enhanced metal-poor stars as
possible, building a sample of $\sim 125$ stars belonging to the rare $r$-II
class.
%
The RPA is envisioned as a multi-stage, multi-year effort to provide
observational, theoretical, and laboratory-based constraints on the nature and origin of
the astrophysical $r$-process. Even at this early stage, this effort has already
identified two $r$-II stars with detected thorium and uranium
\citep{placco2017,holmbeck2018}, a bright $r$-II star at \metal$\sim -2$
\citep{sakari2018}, the first CEMP-$r+s$ star \citep{gull2018}, and a metal-poor
star ([Fe/H] $= -1.47$) with an extreme $r$-process enhancement
\citep{roederer2018}. In addition, the RPA has generated four catalogs of stars
of particular interest: one with candidates selected from medium-resolution
spectroscopy from RAVE \citep[RAdial Velocity Experiment;][]{steinmetz2006,kordopatis2013},
published by \citet{placco2018} and three
with high-resolution spectroscopic follow-up observations (\citealt{hansen2018};
\citealt{sakari2018b}; Ezzeddine et al. 2018 - in prep.). The present
paper is the fifth RPA catalog.

The Best \& Brightest Survey \citep[B\&B;][]{schlaufman2014} made use of
the contrast in the mid-IR photometric bands from the {\emph{WISE}}
\citep[Wide-field Infrared Survey Explore;][]{wright2010} satellite mission to
ground-based optical and near-IR photometry to select over 11,000 candidate VMP
and EMP stars, with an overall success rate of 30\% (VMP) and $\sim 5$\% (EMP),
which is competitive with previous surveys \citep[see,
e.g.,][]{schorck2009,youakim2017}. 
High-resolution spectroscopic follow-up of selected candidates successfully
identified some of the first metal-poor stars in the inner Galaxy
\citep{casey2015} and also the most neutron-capture poor star ever observed
\citep{casey2017b}.
The B\&B survey has the
advantage that all of their candidates are brighter than $V=14.0$, where many other
surveys saturate. By obtaining medium-resolution spectroscopy of B\&B
candidates, we have the opportunity to assemble a definitive sample of relatively bright
CEMP and non-CEMP stars at low metallicity, enabling studies of the known
sub-classes of CEMP stars (CEMP-$s$, CEMP-$r$, CEMP-$i$, CEMP-no), as well as
to identify $r$-I and $r$-II candidates, both of which are ideal for future
high-resolution spectroscopic observations from the ground and in space
\citep[see, e.g., ][]{roederer2012,roederer2012d,roederer2014c,placco2014b,placco2015b,roederer2016}.

\begin{figure*}[!ht]
\epsscale{2.35}
\plottwo{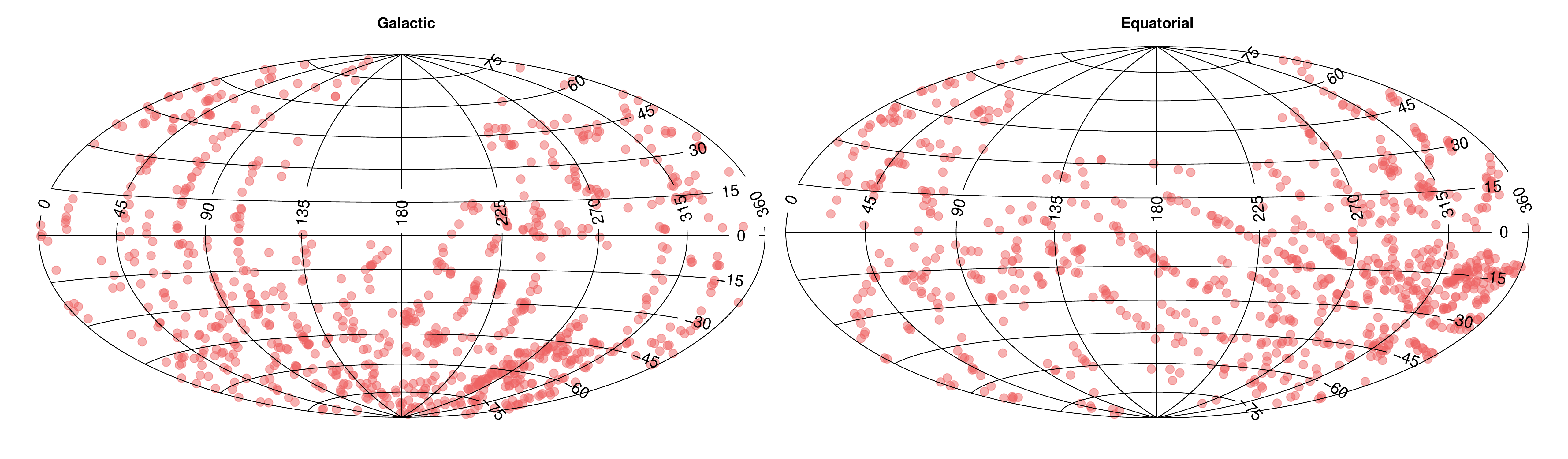}{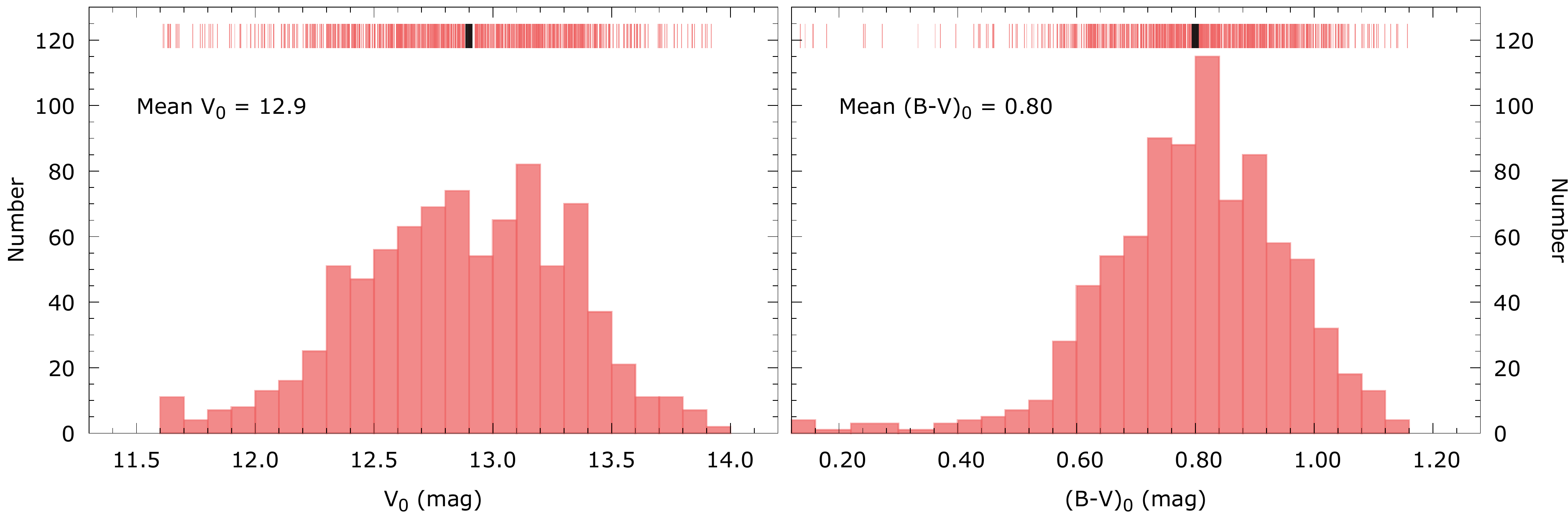}
\caption{Upper panels: Galactic and equatorial coordinates for the observed
targets. Lower panels: Distributions of absorption-corrected $V_0$ magnitudes
and de-reddened $(B-V)_0$ colors. Stripe-density profiles are shown above the
histograms, with the average value highlighted in black.}
\label{coords}
\end{figure*}

This paper reports on the medium-resolution ($R \sim 1,200 - 2,000$) spectroscopic
follow-up of low-metallicity star candidates selected from the B\&B survey.  The
main goal is to determine atmospheric parameters and carbon abundances for a
large sample of stars, which will be used as criteria for targeted
high-resolution spectroscopic follow-ups, including the RPA.
This paper is outlined as follows: Section~\ref{secobs} describes the
medium-resolution spectroscopic observations, followed by the estimates of
the stellar atmospheric parameters and abundances in Section~\ref{secatm}.  We
describe the main abundance trends of our targets, based on \cfe\ and \afe, in
Section~\ref{absec}, and compare our stellar parameter determinations with results
obtained from high-resolution spectroscopic observations by the RPA in
Section~\ref{secrave}.  Our conclusions and prospects for future work are
provided in Section~\ref{final}.

\section{Target Selection and Observations}
\label{secobs}

The medium-resolution spectroscopic follow-up campaign was conducted from semesters 
2015B to 2017A, and collected spectra of 857 unique metal-poor candidates. Below
we describe the target selection from the B\&B database and the subsequent
observations.

\subsection{Target Selection from the B\&B Database}

The candidates for the spectroscopic follow-up were selected from two different
versions of the B\&B database, $v1$ and $v2$, the latter having an improved
selection criteria. Observations conducted in semester 2015B made use of $v1$,
restricted in magnitude by $V > 13.2$ (in order to avoid conflict with a similar
survey of brighter B\&B stars we were already conducting). For semesters 2016A,
2016B, and 2017A, $v2$ was used, with the following restrictions: 

\begin{itemize}
\itemsep-0.25em 

\item[$\star$] Magnitude: $12.5 < V < 13.2$, 
\item[$\star$] Proper motion: $ 0 <$ $\mu_{\rm total}$\footnote{$\mu_{\rm total}
= \sqrt{\mu_{\rm R.A.}^2 + \mu_{\rm decl.}^2}$. Proper motions retrieved from
the fourth US Naval Observatory CCD Astrograph Catalog
\citep[UCAC4;][]{zacharias2013}.} (mas~yr$^{-1}$) $< 25$,
\item[$\star$] Reddening: $E(B-V) < 0.05$.

\end{itemize}

\begin{figure*}[!ht]
\epsscale{1.15}
\plotone{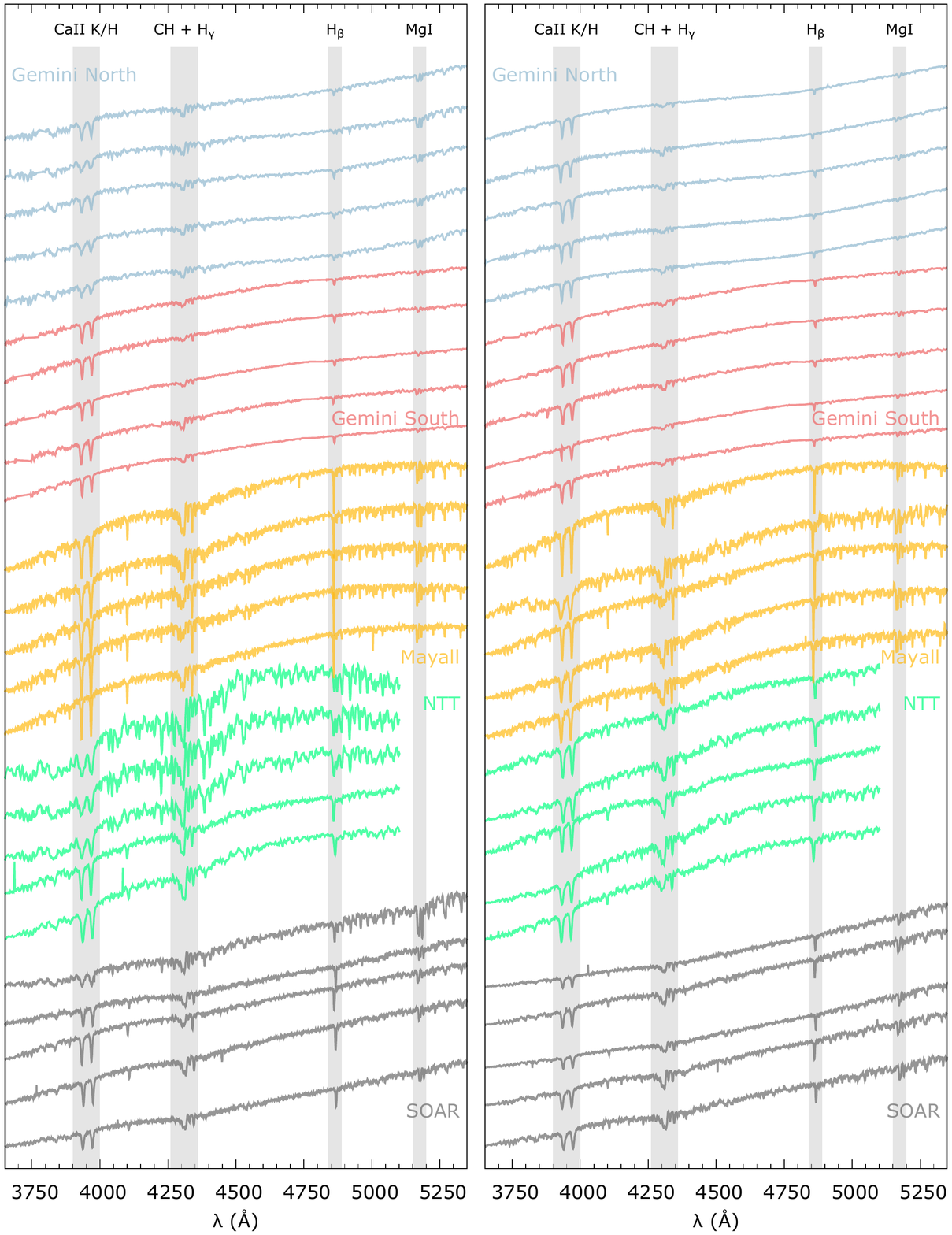}
\caption{Example spectra for 50 (randomly selected) program stars. The colors represent the five
different telescopes used for the observations. The shaded areas highlight
regions of interest (\ion{Ca}{2}, CH G-band$+$H$\gamma$, H$\beta$, and
\ion{Mg}{1}, respectively).}
\label{specex}
\end{figure*}

In total, 71 stars were observed from $v1$ and 786 from $v2$. Stars in the $V <
12.5$ magnitude range were observed by a different program and the restrictions
in proper motion / reddening were designed to minimize contamination from the
numerous foreground disk-like stars. The effectiveness of these criteria in the
selection of metal-poor stars is further evaluated in Section~\ref{secatm}
below.

Table~\ref{telescope} lists the object name, observation date, telescope,
instrument, program ID, and exposure time for the observed candidates;
Table~\ref{coordsmags} lists their coordinates, magnitudes, color indices, and
reddening estimates from the dust maps of \citet{schlafly2011}. 
$J$ and $K$ magnitudes (with photometric quality flags) were
retrieved from the Two Micron All Sky Survey \citep[2MASS;][]{skrutskie2006},
while $B$ and $V$ magnitudes from the AAVSO Photometric All-Sky Survey
\citep[APASS;][]{henden2014}.
The upper panels of Figure~\ref{coords} show the Galactic (left) and equatorial
(right) coordinates of the observed targets and the lower panels show the
distribution of their extinction-corrected $V_0$ magnitudes (left) and
de-reddened $(B-V)_0$ color indices (right). Also shown are the stripe-density
profiles and the average values. By design, this sample consists of mostly
bright ($V < 13.2$) and cool (\teff$ < 5500$~K) low-metallicity candidates,
which are ideal for the high-resolution spectroscopic follow-up program being
executed by the RPA.

\begin{figure*}[!ht]
\epsscale{1.15}
\plotone{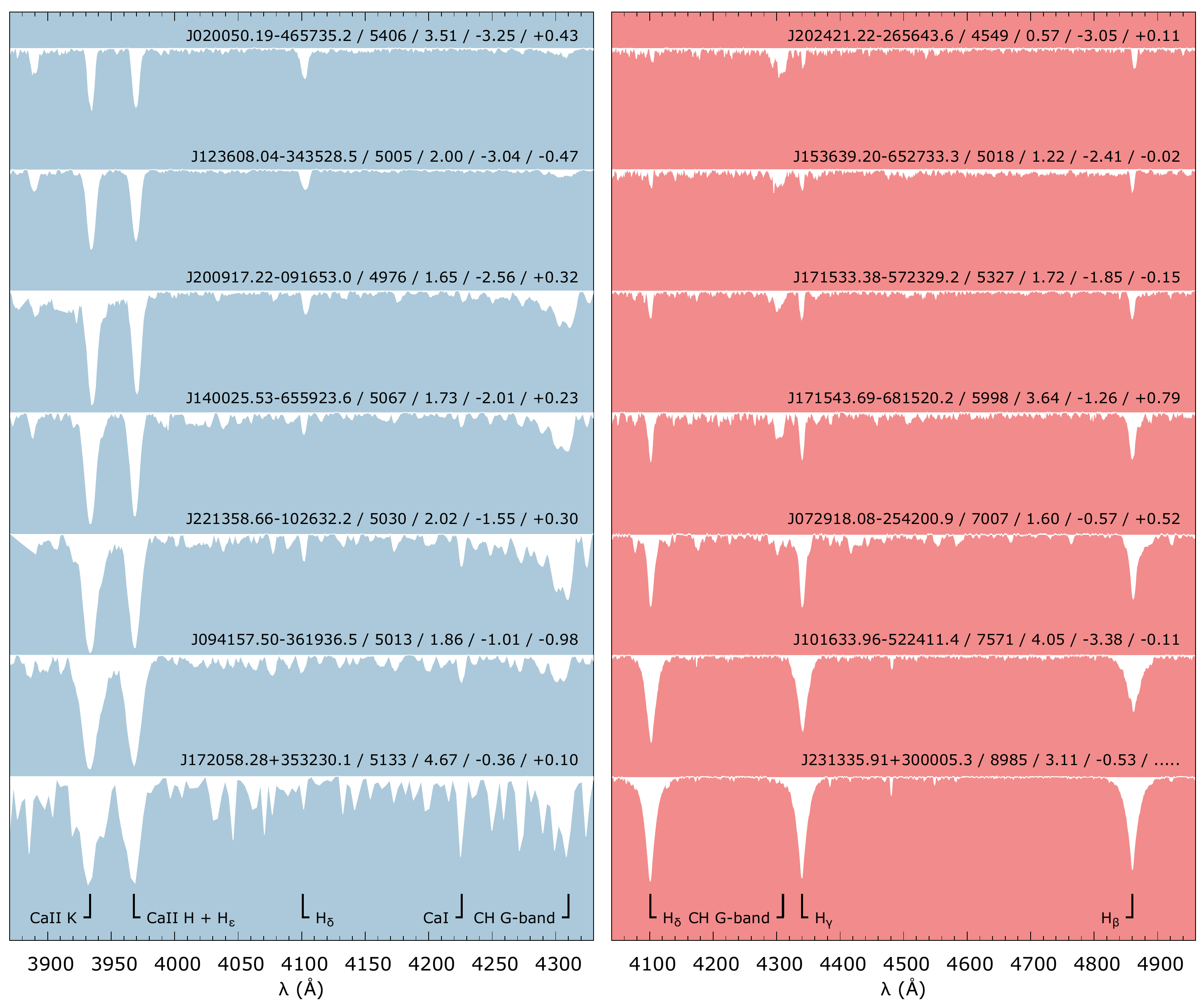}
\caption{Left panel: Observed spectra with similar temperatures, showing the
effect of increasing metallicities on the strength of the \ion{Ca}{2} absorption
features. Right panel: Effect of increasing temperatures on the hydrogen Balmer
absorption features. Listed for each spectrum are:
NAME / \protect\teff\,/ \protect\logg\,/ \protect\metal\,/ \protect\cfe.}
\label{medspec}
\end{figure*}

\subsection{Medium-Resolution Follow-up Observations}
\label{subobs}


Spectroscopic data were gathered using five telescope/instrument setups: 
(i)   SOAR/Goodman, 
(ii)  Gemini North/GMOS-N,
(iii) Gemini South/GMOS-S,
(iv)  Mayall/KOSMOS, and
(v)   NTT/EFOSC-2.
For consistency across the different instruments, we chose grating/slit
combinations that would yield a resolving power $R\sim 1,200-2,000$, and
exposure times sufficient to reach a signal-to-noise ratio of at least S/N$\sim
30$ per pixel at the \ion{Ca}{2}~K line (3933.3\,{\AA}). Calibration frames
included arc-lamp exposures, bias frames, and quartz flats. All tasks related to
spectral reduction, extraction, and wavelength calibration were performed using
standard IRAF\footnote{\href{http://iraf.noao.edu}{http://iraf.noao.edu}.}
packages.

Figure~\ref{specex} shows the spectra for 50 stars (randomly selected from the
857 star database) followed-up in this work, color-coded by the different
telescopes used for the observations. The shaded areas highlight wavelength
regions of interest for atmospheric parameter and abundance determinations (see
Section~\ref{secatm} for details). Even though there are noticeable differences
in coverage, CCD response, and resolution, these spectra are all within acceptable
ranges for the analysis performed in this work.  Details on each observing setup
are provided below. 

\paragraph{Gemini North and South Telescopes}

474 stars were observed with the twin 8.1\,m Gemini North (134
stars) and Gemini South (340 stars) telescopes and the
GMOS \citep[Gemini Multi-Object Spectrographs;][]{davies1997,gimeno2016} instruments.
In both cases, we used the
B600~l~mm$^{\rm{-1}}$ grating (G5323 for GMOS South and G5307 for GMOS North)
and a 1$\farcs$0 slit, 
resulting in a wavelength coverage in the range [3200:5800]\,{\AA} at resolving power $R \sim 2,000$. 
%

\paragraph{ESO New Technology Telescope}

256 stars were observed with the 3.58\,m New Technology Telescope
(NTT), located at La Silla Observatory, part of the European Southern
Observatory.
We used the EFOSC-2 \citep[ESO Faint Object Spectrograph and Camera v.2;][]{buzzoni1984} 
instrument with Grism\#7 (600~gr~mm$^{\rm{-1}}$) and a 1$\farcs$0 slit, 
resulting in a wavelength coverage in the range [3300:5100]\,{\AA} at resolving power $R \sim 1,200$. 
%

\paragraph{KPNO Mayall Telescope}

73 stars were observed with the 4\,m Mayall telescope, located at
Kitt Peak National Observatory, using the KOSMOS 
\citep[Kitt Peak Ohio State Multi-Object Spectrograph;][]{martini2014}
instrument.
We used the 600~l~mm$^{\rm{-1}}$ grating, the blue
setting, and a 0$\farcs$9 slit, 
resulting in a wavelength coverage in the range [3600:6300]\,{\AA} at resolving power $R \sim 1,800$. 
%

\paragraph{SOAR Telescope}

54 stars were observed with the 4.1\,m Southern Astrophysical
Research (SOAR) telescope. 
The Goodman Spectrograph \citep{clemens2004} was used with the
600~l~mm$^{\rm{-1}}$ grating, the blue setting, and a 1$\farcs$0 slit,
resulting in a wavelength coverage in the range [3600:6200]\,{\AA} at resolving power $R \sim 1,500$. 
%

\begin{figure}[!ht]
\epsscale{1.15}
\plotone{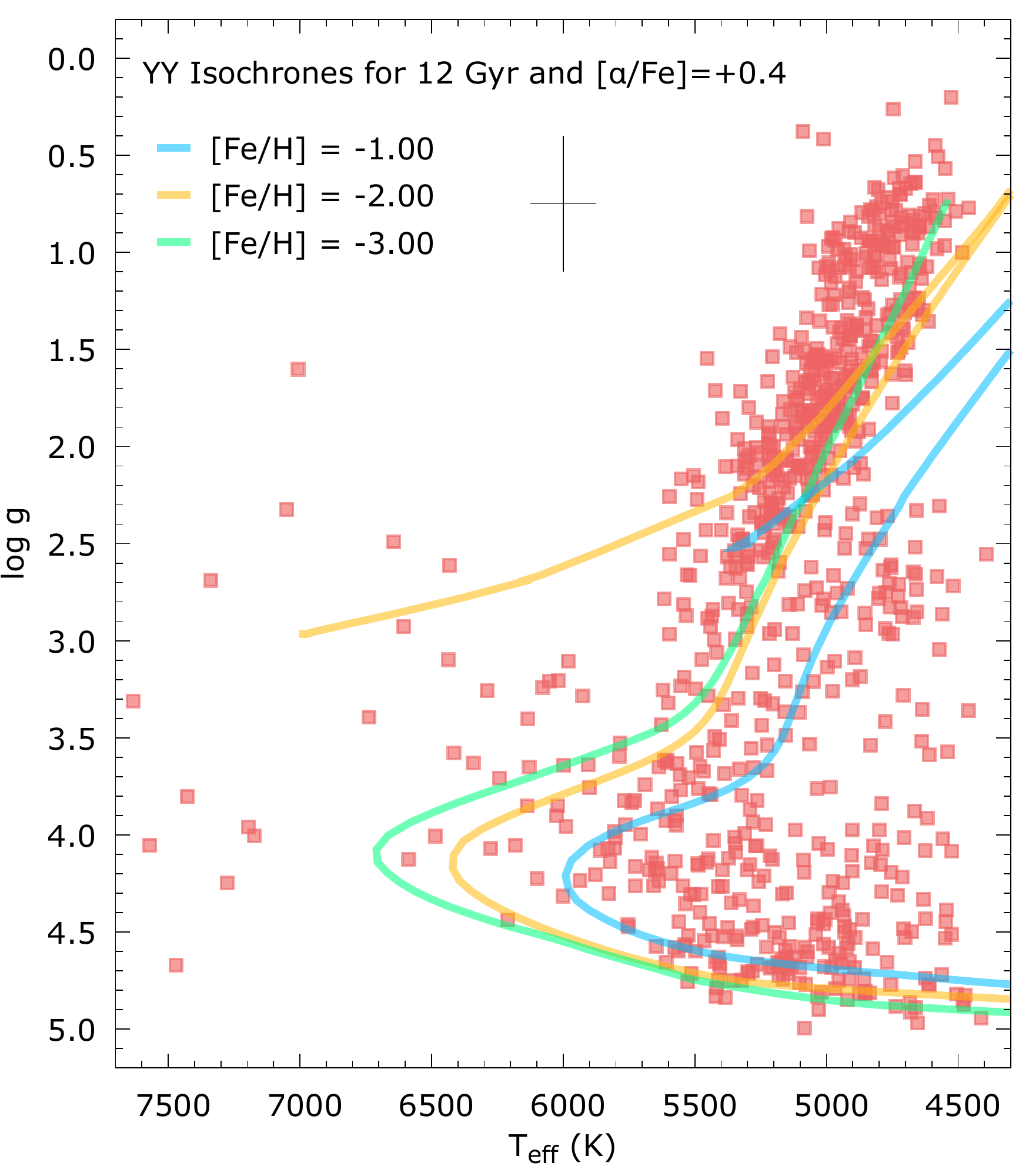}
\caption{Surface gravity vs. \teff\ (H-R) diagram for the program stars, using the parameters calculated by
the n-SSPP, listed in Table~\ref{comprave}. Overplotted are the YY Isochrones
\citep[12~Gyr, 0.8~M$_{\rm \odot}$, \afe=$+$0.4;][]{demarque2004} for
\metal\ = $-$2.0, $-$2.5, and $-$3.0, and horizontal-branch tracks from
\citet{dotter2008}.}
\label{isochrone}
\end{figure}

\section{Stellar Parameters and Abundances}
\label{secatm}

Stellar atmospheric parameters (\teff, \logg, and \metal), as well as carbonicity (\cfe)
and $\alpha$-to-iron ratios (\afe), were determined using the n-SSPP
\citep{beers2014,beers2017}, a modified version of the SEGUE Stellar Parameter
Pipeline \citep[SSPP;][]{lee2008a,lee2008b,lee2011,lee2013}.
The n-SSPP uses as input the observed spectrum and photometric information for a
given object. There are several internal routines that calculate the atmospheric
parameters based on spectral line indices, photometric calibrations, and
matching with a database of synthetic spectra.
The \cfe\, and \afe\, are estimated from the strength of the CH G-band molecular
feature at $\sim 4300$\,{\AA} and the \ion{Mg}{1} triplet at 5150-5200\,{\AA},
respectively.
Details on the n-SSPP processing for spectra similar to the ones analyzed in
this work can be found in \citet{placco2018}.

Figure~\ref{medspec} shows, in the left column of panels, the effect of changes in
metallicity on the \ion{Ca}{2} spectral lines for stars with similar \teff, and
increasing \metal\, (from \metal\ =\,$-3.25$ to \metal\ =\,$-0.36$).  Listed for each
spectra are: NAME / \teff\,/ \logg\,/ \metal\,/ \cfe.  At this resolving power, the
\ion{Ca}{2}~K line is the main proxy for metallicity in the optical wavelength
regime.  The right column of panels in Figure~\ref{medspec} shows spectra with increasing
\teff\, (from \teff\ = 4549~K to \teff\ = 8985~K) and its effect on the strength of
three hydrogen Balmer lines.

The n-SSPP was able to estimate \teff\ and \logg\ for 842 out of the 857
stars observed. The 15 stars without adopted parameters had low S/N spectra and/or
large mismatches between the color-based temperatures and the spectroscopic
calibrations.
Metallicities were determined for $\sim 93\%$ of the observed sample (796
stars). The non-determinations arise from lack of temperature estimates by the
n-SSPP, or stars with core emission in the \ion{Ca}{2}~K line.  The \cfe\ and \afe\
abundance ratios were estimated for 793 and 584 stars, respectively.  The
carbon-abundance determination is not carried out for low-quality spectra
(mostly S/N $<$ 10), or in spectra where the CH $G$-band molecular feature is too
weak to be reliably distinguished from the underlying noise \citep[usually for
\teff\ $ > 6500$~K; see][]{placco2016}.  In addition, due to the lack of spectral
coverage, \afe\ was not obtained for the NTT/EFOSC-2 spectra.
Final atmospheric parameters and abundances for the sample are listed in
Table~\ref{comprave}. 
Also included in the table are the corrections for carbon abundances, based on
the stellar-evolution models presented in \citet{placco2014c}, the final \cfe,
and $A$(C)\footnote{$A$(C) = $\log(N_C/{}N_H) + 12$}, the latter two including
the corrections.  

\begin{figure*}[!ht]
\epsscale{1.15}
\plottwo{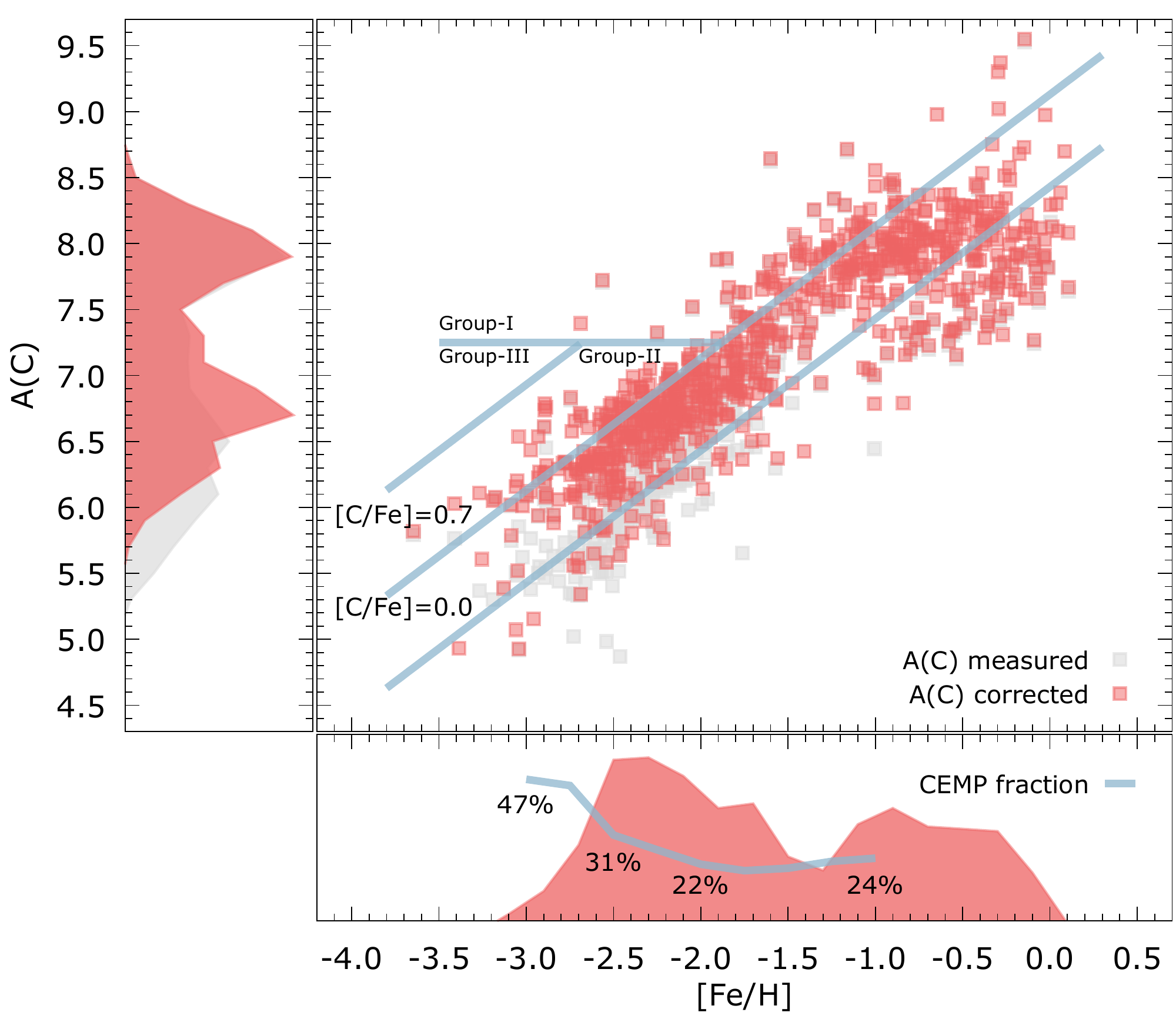}{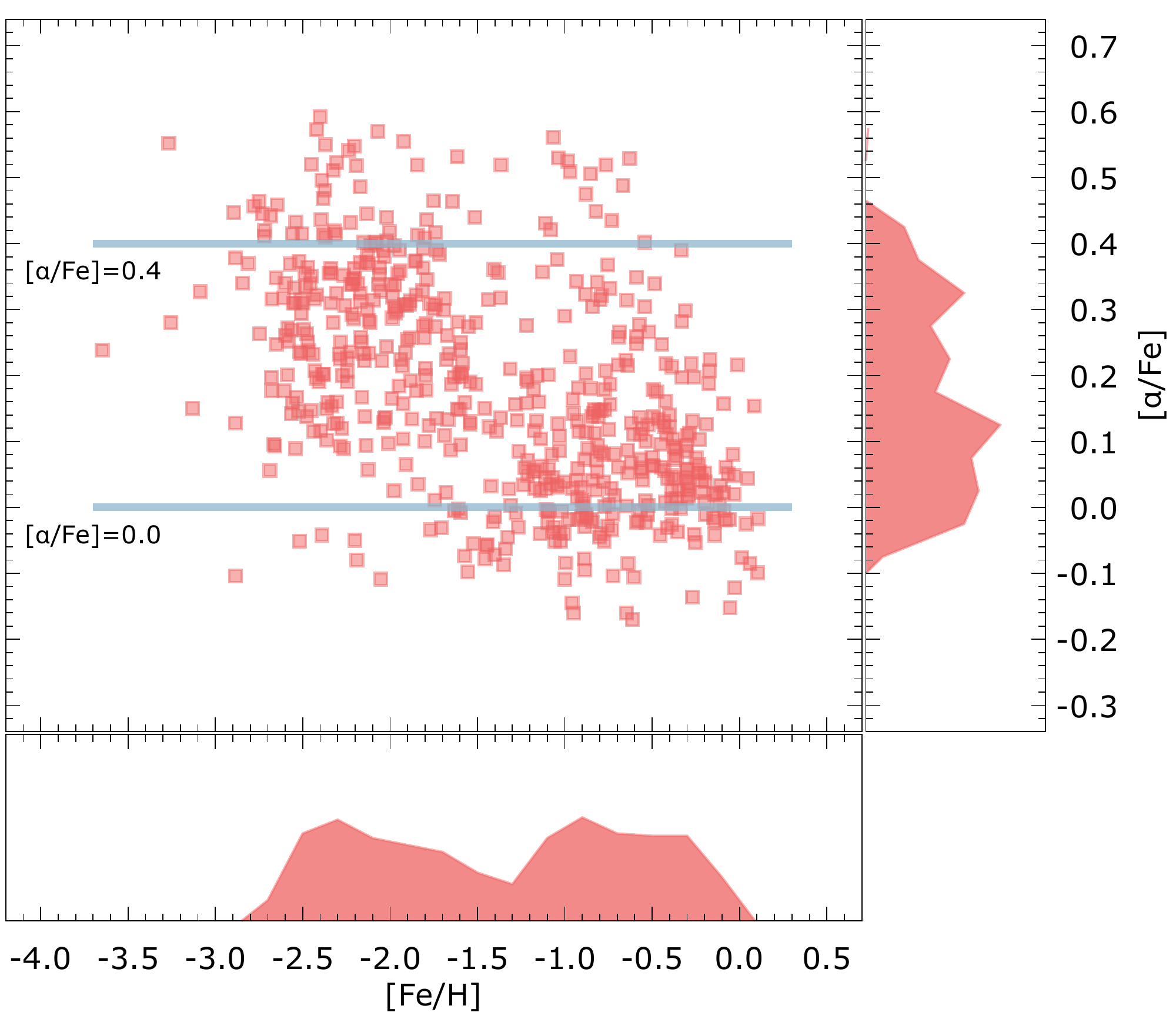}
\caption{Absolute carbon, ($A$(C), corrected as described in the text - left panel), and
$\alpha$-element abundance ratios, [$\alpha$/Fe] (right panel), as a function of the metallicity
calculated by the n-SSPP. The side and lower panels show the marginal
distributions for each quantity. The solid line in the lower panel shows the
cumulative CEMP fractions for the stars with $-3.0 \leq$\,\metal\,$\leq -1.0$.}
\label{acfeh} 
\end{figure*}

With the calculated metallicities, we were able to test the effectiveness of the
proper motion and reddening cuts described in Section~\ref{secobs}. From the
$v1$ database, for which no extra proper motion and reddening cuts were applied,
48\% of the observed targets have \metal\ $\leq -1.0$ and 9\% have \metal\ $\leq
-2.0$.  These numbers increase to 71\% for \metal\ $\leq -1.0$ and 42\% for
\metal\ $\leq -2.0$, when observing candidates selected from the $v2$ database,
with the cuts applied. By combining both sub-samples, 69\% of the stars have
\metal\ $\leq -1.0$, 39\% have \metal\ $\leq -2.0$, and 2\% have \metal\ $\leq
-3.0$. These fractions are somewhat smaller than the values reported by
\citet{placco2018} for the follow-up of RAVE low-metallicity candidates.
However, the RAVE stars already had preliminary \metal\ estimates from their
moderate-resolution spectra, while the B\&B star candidates were originally
selected based on photometry alone.

The distribution of effective temperatures and surface gravities derived for the
B\&B follow-up sample is shown in Figure~\ref{isochrone}. Solid lines represent
the Yale-Yonsei Isochrones \citep[12~Gyr, 0.8~M$_{\rm \odot}$,
\afe\ = $+$0.4;][]{demarque2004} for \metal\ = $-$2.0, $-$2.5, and $-$3.0. Also
shown are the Horizontal-Branch tracks from the Dartmouth Stellar Evolution
Database \citep{dotter2008}, for the same input parameters. Since the B\&B is a
magnitude-limited survey, and further brightness constraints were applied before
observations, it is expected that the present sample would be dominated by stars
in the sub-giant and giant evolutionary stages.
There is a noticeable shift of about $+150$~K between the data and the
isochrones. The same offset is seen in the RAVE stars followed-up in
\citet{placco2018}, even though the \teff\, values for that sample agree well
with estimates based on the infrared flux method of \citet{casagrande2010}.
Comparison with parameters determined from high-resolution spectroscopy within
the RPA will help address these differences.
Further details are provided in Section~\ref{secrave}.
Typical uncertainties for the atmospheric parameters calculated by the n-SSPP
are 125\,K for \teff, 0.35\,dex for \logg, and 0.15-0.20\,dex for \metal, \cfe,\
and \afe.

\section{Carbonicity and $\alpha$-to-iron \\ abundance ratios}
\label{absec}

The carbonicity and $\alpha$-to-iron ratios estimated by the n-SSPP can provide useful
constraints on the formation scenarios for metal-poor stars, and also serve as
criteria to assemble lists for high-resolution spectroscopic follow-up.
Figure~\ref{acfeh} shows the distribution of the carbon abundances ($A$(C), as
corrected - left side) and the $\alpha$-element abundance ratios (right side), as a
function of the \metal\ estimated in this work.  The lower and side panels show
marginal distributions for each quantity.  
There is no significant trend for the \afe\ ratios, with values ranging from
$-0.2$ to $+0.6$. These are within expectation for samples with similar \metal\
ranges and Galactic chemical evolution models \citep{reggiani2017}.

The $A$(C) vs. \metal\ diagram provides an important diagnostic for the
type(s) of progenitor(s) that could have formed a given star
\citep{spite2013,bonifacio2015,hansen2015,yoon2016}. \citealt{yoon2016} proposed
a classification of CEMP stars, based on $A$(C) vs. [Fe/H] (both of which can be
estimated using medium-resolution spectra alone), into three groups, the
so-called Yoon-Beers diagram.  Using the criteria described in
\citet{placco2018}, shown in Figure~\ref{acfeh}, we find 97 stars in Group~I and
36 stars in Group~II, with no stars belonging to Group~III. The stars in
Group~II, which are likely to be CEMP-no (\cfe\ $ \geq +0.7$ and \xfe{Ba}\ $ <
0.0$), are ideal targets for high-resolution spectroscopic follow-up and
determination of their light-element chemical abundance patterns, as such stars
provide precious information about nucleosynthesis pathways in the early
universe.

Since the original B\&B sample did not have any indicators of carbon
enhancement, one would expect that the CEMP fractions for the stars observed in
this work are similar to values from the literature for other ``carbon-blind''
samples.  The lower left panel of Figure~\ref{acfeh} shows the CEMP fractions
(using the corrected \cfe,\ values - see Section~\ref{secatm} for further
details) for metallicities in the range $[-3.0,-1.0]$. 
There is an overall good agreement (within 1-$\sigma$) with the values from
\citet{placco2018}. 
The fraction found for
\metal$ < -2.5$ in this work ($31^{+9}_{-7}\%$\footnote{Uncertainties in the
fractions are represented by the Wilson score confidence intervals
\citep{wilson1927}.}) agrees very well with the fraction found in
\citet{schlaufman2014} for the same metallicity range, $28^{+18}_{-13}\%$.
There is also a good agreement for the fractions at
\metal$\leq -2.0$ and $-3.0$ from this work ($22^{+5}_{-4}\%$ and
$47^{+22}_{-21}\%$, respectively) and the fractions calculated using abundances
from high-resolution spectroscopy \citep[20\pc\, and 43\pc;][]{placco2014c}.



\begin{figure*}[!ht]
\epsscale{1.15}
\plotone{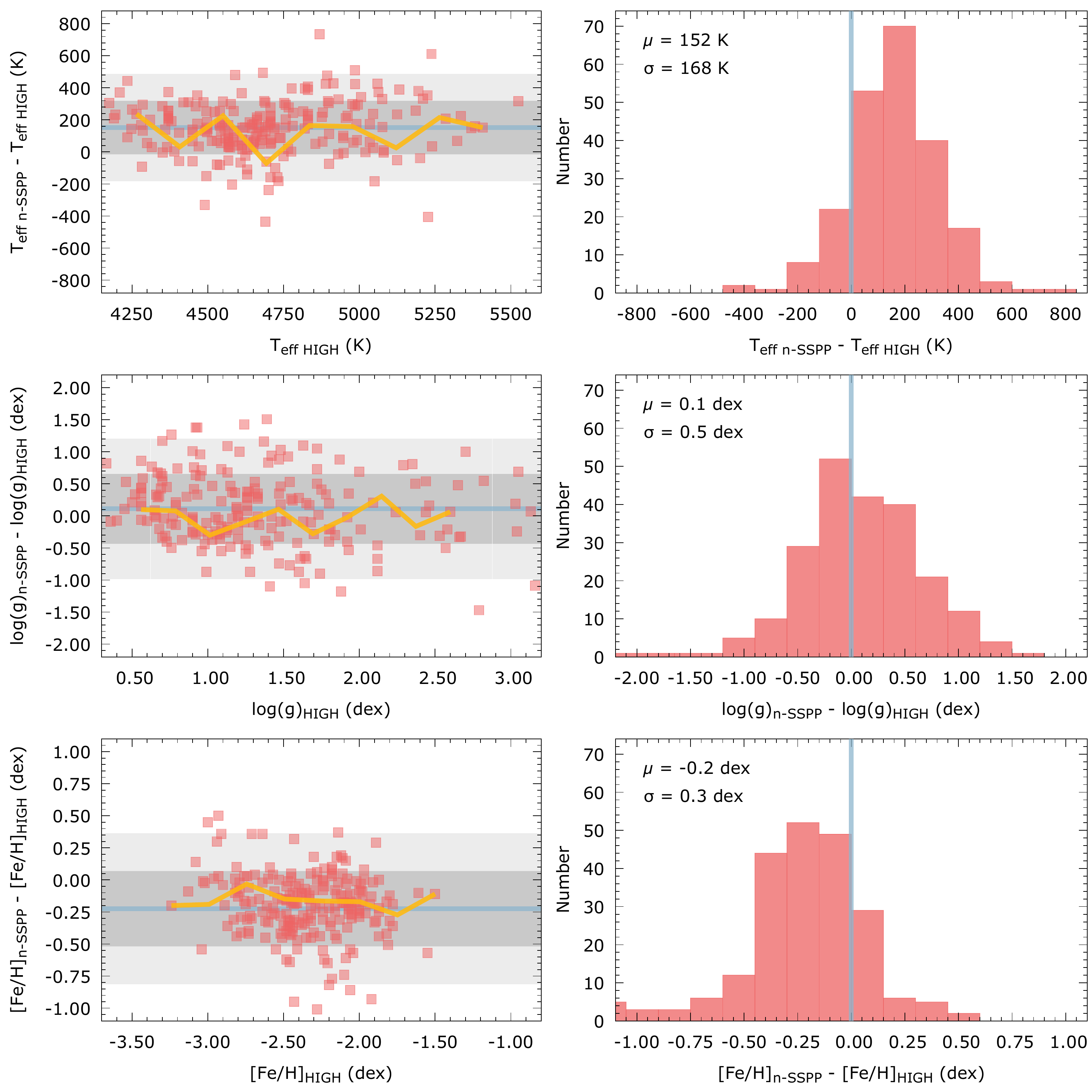}
\caption{Left panels: Differences between the parameters determined
by the n-SSPP, $T_{\rm eff\, n-SSPP}$, $\log\,g_{\rm\, n-SSPP}$, and [Fe/H]$_{\rm\,
n-SSPP}$, and the values from high-resolution, $T_{\rm eff\, HIGH}$,
$\log\,g_{\rm\, HIGH}$, and [Fe/H]$_{\rm\, HIGH}$, as a
function of the high-resolution spectroscopic values.
The horizontal solid line is the average of the residuals, while the
darker and lighter shaded areas represent the 1-$\sigma$ and 2-$\sigma$ regions,
respectively. Also shown are locally weighted regression ({\emph{loess}}) lines.
Right panels: Histograms of the residuals between the 
n-SSPP and high-resolution parameters shown in the left panels. Each panel also
lists the average and $\sigma$ of the residual distribution.}
\label{highcomp}
\end{figure*}


\section{Comparison with high-resolution spectroscopic data from the RPA}
\label{secrave}

In this section we present a comparison between the atmospheric parameters
estimated in this work and values\footnote{For this comparison we used the
parameters calculated assuming local thermodynamic equilibrium.} from the RPA
high-resolution data release papers \citep{hansen2018,sakari2018b}. We combine
the medium-resolution spectra from this work with data from \citet{placco2018},
which are of similar resolution, and were also processed by the n-SSPP. In
total, 218 stars were used for these comparisons; results are presented in
Figure~\ref{highcomp}.
The left column of panels shows the differences between the parameters
determined by the n-SSPP, $T_{\rm eff\, n-SSPP}$, $\log\,g_{\rm\, n-SSPP}$, and
[Fe/H]$_{\rm\, n-SSPP}$, and the values from high-resolution, $T_{\rm eff\,
HIGH}$, $\log\,g_{\rm\, HIGH}$, and [Fe/H]$_{\rm\, HIGH}$, as a function of the
high-resolution spectroscopic values.
The horizontal solid line is the average of the residuals, while the darker and
lighter shaded areas represent the 1-$\sigma$ and 2-$\sigma$ regions,
respectively. Also shown are locally weighted regression ({\emph{loess}}) lines.
The right column of panels shows histograms of the residuals between the n-SSPP
and high-resolution parameters. Each panel also lists the average and standard
deviation of the residual distribution.

Inspection of Figure~\ref{highcomp}, in particular the {\emph{loess}} lines,
reveals that there are no relevant trends in the comparisons other than constant
shifts (values from this work minus parameters from the RPA) for \teff\,
(152\,K), \logg\, (0.10\,dex), and \metal\, ($-0.2$\,dex). 
In addition, the
standard deviations of the residuals (168\,K for \teff, 0.5\,dex for \logg, and
0.3\,dex for \metal), are within the expected values for such comparisons
\citep[e.g.,][]{lee2013,beers2014}. 
It is worth noting that the shifts \teff\, and \metal\, are not independent,
since changes in temperature affect the strength of the
absorption features in the stellar atmospheres.
Further analyses and corrections for the
medium-resolution parameters will be conducted once the first phase of
``snapshot'' (moderate S/N, moderately high-resolution) observations of the RPA
are finished, which should yield a database of at least 2,000 stars with
high-resolution determinations for comparison.

\section{Conclusions}
\label{final}

We have presented results from a medium-resolution ($R \sim 1,200 - 2,000$)
spectroscopic follow-up of low-metallicity star candidates selected from the
Best \& Brightest Survey.  Our observing campaign ran from semesters 2015B to
2017A, and used five different telescope/instrument configurations, in both the
Southern and Northern Hemispheres.  Atmospheric parameters and abundances for
carbon and the $\alpha$-elements were calculated using our well-tested n-SSPP
pipeline. From the 857 unique stars observed, 553 were confirmed to be
metal-poor (\metal\ $\leq -1.0$), and 133 were carbon-enhanced (\cfe\ $ \geq
+0.7$), after evolutionary corrections have been applied. There are 36 CEMP
Group~II stars that are currently being followed-up in high-resolution, to
determine their chemical abundance patterns, and compare with yields from
theoretical models of Population III stellar nucleosynthesis. We also showed
that the success rate for the identification of very low-metallicity stars can
be significantly increased (from 9\pc,\ to 42\pc,\ for \metal\ $ \leq -2.0$),
when proper motions are also used as a selection criteria, primarily due to the
exclusion of foreground disk-like stars.

Comparisons between the parameters determined in this work and values from the
RPA catalogs reveal that the residual zero-point offsets are within 1-$\sigma$
for \teff,\ (152~K), \logg,\ (0.10~dex), and \metal,\ ($-0.2$~dex), which allow
for a successful target selection for high-resolution spectroscopic follow-up.
The catalog generated by this work will continue to serve as a reliable source
of targets for the RPA and other projects in the future.


\acknowledgments
We thank Kevin C. Schlaufman and Andrew R. Casey for kindly providing
early access to the candidate lists from the Best \& Brightest Survey.
The authors acknowledge partial support for this work from grant PHY 1430152;
Physics Frontier Center/JINA Center for the Evolution of the Elements
(JINA-CEE), awarded by the US National Science Foundation (NSF).  
J.C. acknowledges support from CONICYT project Basal AFB-170002 and by the
Chilean Ministry for the Economy, Development, and Tourisms Programa Iniciativa
Científica Milenio grant IC 120009, awarded to the Millennium Institute of
Astrophysics.
S.R.\ acknowledges partial financial support from FAPESP and CNPq. This study
was financed in part by the Coordena\c{c}\~{a}o de Aperfei\c{c}oamento de
Pessoal de N\'{i}vel Superior - Brasil (CAPES) - Finance Code 001.
Y.S.L.\ acknowledges support from the National Research Foundation (NFR) of
Korea grant funded by the Ministry of Science and ICT (No.2017R1A5A1070354 and
NRF-2018R1A2B6003961).
E.S.\ and K.Y.\ gratefully acknowledge funding by the Emmy Noether program from
the Deutsche Forschungsgemeinschaft (DFG).
A.P.J.\ is supported by NASA through Hubble Fellowship grant HST-HF2-51393.001
awarded by the Space Telescope Science Institute, which is operated by the
Association of Universities for Research in Astronomy, Inc., for NASA, under
contract NAS5-26555.
I.U.R.\ acknowledges support from NSF grants AST 1613536 and AST 1815403.
%
%
Based on observations at Kitt Peak National Observatory, National Optical
Astronomy Observatory (NOAO Prop. ID: 
17A-0295), 
which is operated by the Association of Universities for Research in Astronomy
(AURA) under cooperative agreement with the National Science Foundation. The
authors are honored to be permitted to conduct astronomical research on Iolkam
Du'ag (Kitt Peak), a mountain with particular significance to the Tohono
O'odham. 
%
%
Based on observations obtained at the Gemini Observatory (Prop. IDs:
GN-2015B-Q-100, GN-2016B-Q-85, GS-2015B-Q-104, GS-2016A-Q-107, GS-2016B-Q-86), 
which is operated by the Association of Universities for Research in Astronomy,
Inc., under a cooperative agreement with the NSF on behalf of the Gemini
partnership: the National Science Foundation (United States), the National
Research Council (Canada), CONICYT (Chile), Ministerio de Ciencia,
Tecnolog\'{i}a e Innovaci\'{o}n Productiva (Argentina), and Minist\'{e}rio da
Ci\^{e}ncia, Tecnologia e Inova\c{c}\~{a}o (Brazil).
%
%
Based on observations obtained at the Southern Astrophysical Research (SOAR)
telescope (Prop. ID:
2017A-0016), 
which is a joint project of the Minist\'{e}rio da Ci\^{e}ncia, Tecnologia, e
Inova\c{c}\~{a}o (MCTI) da Rep\'{u}blica Federativa do Brasil, the U.S.
National Optical Astronomy Observatory (NOAO), the University of North Carolina
at Chapel Hill (UNC), and Michigan State University (MSU).
%
%
Based on observations collected at the European Organisation for Astronomical
Research in the Southern Hemisphere under ESO programme(s) 
097.D-0196(A), 097.D-0196(B), 098.B-0618(A), 098.D-0434(A), and 099.D-0428(A).
%
%
This research was made possible through the use of the AAVSO Photometric
All-Sky Survey (APASS), funded by the Robert Martin Ayers Sciences Fund.
This publication makes use of data products from the Two Micron All Sky Survey,
which is a joint project of the University of Massachusetts and the Infrared
Processing and Analysis Center/California Institute of Technology, funded by
the National Aeronautics and Space Administration and the National Science
Foundation.
This research has made use of NASA's Astrophysics Data System Bibliographic
Services; the arXiv pre-print server operated by Cornell University; the SIMBAD
database hosted by the Strasbourg Astronomical Data Center; and the online Q\&A
platform {\texttt{stackoverflow}}
(\href{http://stackoverflow.com/}{http://stackoverflow.com/}).

\software{
{\texttt{astrolibR}}\,\citep{astrolibr}, 
{\texttt{AitoffR}}\,\citep{aitoffr}, 
{\texttt{awk}}\,\citep{awk}, 
{\texttt{gnuplot}}\,\citep{gnuplot}, 
{\texttt{IRAF}}\,\citep{tody1986,tody1993}, 
{\texttt{n-SSPP}}\,\citep{beers2014}, 
{\texttt{R-project}}\,\citep{rproject}, 
{\texttt{sed}}\,\citep{sed}.
}


\bibliographystyle{apj}
\bibliography{placco}

\clearpage
\startlongtable

\begin{deluxetable*}{@{}cclllr}
\tabletypesize{\scriptsize}
\tabletypesize{\tiny}
\tablewidth{0pc}
\tablecaption{Observing Details \label{telescope}}
\tablehead{
\colhead{Star Name}   &
\colhead{Date}        &
\colhead{Telescope}   &
\colhead{Instrument}  &
\colhead{Proposal ID} &
\colhead{Exp.}        \\
\colhead{(2MASS)}      &
\colhead{(UTC)}       &
\colhead{}            &
\colhead{}            &
\colhead{}            &
\colhead{(s)}         }
\startdata
\input{table1c.tex}
\enddata
\end{deluxetable*}

\clearpage
\startlongtable

\begin{deluxetable*}{cccrrrrrrcr}
\tabletypesize{\scriptsize}
\tabletypesize{\tiny}
\tablewidth{0pc}
\tablecaption{Coordinates, Magnitudes, Color Indices, and Reddening Estimates \label{coordsmags}}
\tablehead{
\colhead{Star Name} &
\colhead{$\alpha$}  &
\colhead{$\delta$}  &
\colhead{$l$}       &
\colhead{$b$}       &
\colhead{$V$}       &
\colhead{$(B-V)$}   &
\colhead{$J$}       &
\colhead{$(J-K)$}   &
\colhead{ph\_qual\xx}  &
\colhead{$E(B-V)$}  \\
\colhead{(2MASS)}   &
\colhead{(J2000)}   &
\colhead{(J2000)}   &
\colhead{(deg)}     &
\colhead{(deg)}     &
\colhead{(mag)}     &
\colhead{}          &
\colhead{(mag)}     &
\colhead{}          &
\colhead{}          &
\colhead{}          }
\startdata
\input{table2c.tex}
\enddata
\tablenotetext{a}{2MASS JHK photometric quality flag. Further details given in: 
\href{http://vizier.u-strasbg.fr/viz-bin/VizieR-n?-source=METAnot&catid=2246&notid=5}{http://vizier.u-strasbg.fr/viz-bin/VizieR-n?-source=METAnot\&catid=2246\&notid=5}}
\end{deluxetable*}

\clearpage
\startlongtable

\begin{deluxetable*}{@{}ccrrrrrrr}
\tabletypesize{\scriptsize}
\tabletypesize{\tiny}
\tablewidth{0pc}
\tablecaption{Stellar Parameters and Abundances from the n-SSPP \label{comprave}}
\tablehead{
\colhead{Star Name}             &
\colhead{\teff}                 &
\colhead{\logg}                 &
\colhead{\metal}                &
\colhead{\cfe}                  &
\colhead{$\Delta$\cfe\xx}       &
\colhead{\cfe$_{\rm cor}$\yy}   &
\colhead{$A$(C)$_{\rm cor}$\zz} &
\colhead{\afe}                  \\
\colhead{(2MASS)}               &
\colhead{(K)}                   &
\colhead{(cgs)}                 &
\colhead{}                      &
\colhead{}                      &
\colhead{}                      &
\colhead{}                      }
\startdata
\input{table3c.tex}
\enddata
\tablenotetext{a}{Carbon correction from \citet{placco2014c}.}
\tablenotetext{b}{Corrected carbon-to-iron ratio.}
\tablenotetext{c}{Corrected absolute carbon abundance.}
\end{deluxetable*}


\end{document}

%% file: table1c.tex
J000045.79$+$380245.7 &   2015$-$11$-$08 &    Gemini$-$North &     GMOS$-$N &  GN$-$2015B$-$Q$-$100 & 300 \\
J000106.55$+$452812.5 &   2015$-$11$-$08 &    Gemini$-$North &     GMOS$-$N &  GN$-$2015B$-$Q$-$100 & 300 \\
J000111.95$+$032105.0 &   2017$-$07$-$07 &          Mayall &     KOSMOS &        17A$-$0295 & 600 \\
J000123.03$+$495329.1 &   2015$-$11$-$11 &    Gemini$-$North &     GMOS$-$N &  GN$-$2015B$-$Q$-$100 & 370 \\
J000137.83$+$500539.6 &   2015$-$11$-$16 &    Gemini$-$North &     GMOS$-$N &  GN$-$2015B$-$Q$-$100 & 600 \\
J000212.23$-$224138.9 &   2015$-$11$-$09 &    Gemini$-$South &     GMOS$-$S &  GS$-$2015B$-$Q$-$104 & 600 \\
J000216.68$-$245349.5 &   2015$-$11$-$09 &    Gemini$-$South &     GMOS$-$S &  GS$-$2015B$-$Q$-$104 & 600 \\
J000312.65$-$001504.4 &   2015$-$11$-$24 &    Gemini$-$South &     GMOS$-$S &  GS$-$2015B$-$Q$-$104 & 600 \\

%% file: table2c.tex
J000045.79$+$380245.7 & 00:00:45.84 & $+$38:02:45.6 & 112.055 & $-$23.753 &   13.883 &    0.958 &   11.903 &     0.657 & AAA &  0.089 \\
J000106.55$+$452812.5 & 00:01:06.48 & $+$45:28:12.0 & 113.764 & $-$16.504 &   13.348 &    1.042 &   11.317 &     0.722 & AAA &  0.118 \\
J000111.95$+$032105.0 & 00:01:12.00 & $+$03:21:03.6 &  99.309 & $-$57.198 &   12.803 &    0.809 &   11.049 &     0.620 & AAA &  0.021 \\
J000123.03$+$495329.1 & 00:01:23.04 & $+$49:53:27.6 & 114.722 & $-$12.181 &   13.693 &    0.929 &   11.995 &     0.631 & AAA &  0.116 \\
J000137.83$+$500539.6 & 00:01:37.92 & $+$50:05:38.4 & 114.802 & $-$11.990 &   13.803 &    0.916 &   11.993 &     0.666 & AAA &  0.118 \\
J000212.23$-$224138.9 & 00:02:12.24 & $-$22:41:38.4 &  51.912 & $-$77.996 &   13.273 &    0.758 &   11.727 &     0.564 & AAA &  0.018 \\
J000216.68$-$245349.5 & 00:02:16.80 & $-$24:53:49.2 &  41.605 & $-$78.737 &   13.478 &    0.841 &   11.841 &     0.595 & AAA &  0.016 \\
J000312.65$-$001504.4 & 00:03:12.72 & $-$00:15:03.6 &  97.639 & $-$60.733 &   13.641 &    1.148 &   11.298 &     0.754 & AAA &  0.029 \\

%% file: table3c.tex
J000045.79$+$380245.7  &    4748 &     2.97 &  $-$0.35 &  $-$0.59 &  $+$0.03 &  $-$0.56 &  $+$7.52 &  $+$0.03 \\
J000106.55$+$452812.5  & \nodata &  \nodata &  \nodata &  \nodata &  \nodata &  \nodata &  \nodata &  \nodata \\
J000111.95$+$032105.0  &    4984 &     1.89 &  $-$2.05 &  $+$0.14 &  $+$0.13 &  $+$0.27 &  $+$6.65 &  $+$0.22 \\
J000123.03$+$495329.1  &    5280 &     3.93 &  $-$0.69 &  $+$0.30 &     0.00 &  $+$0.30 &  $+$8.04 &  $+$0.26 \\
J000137.83$+$500539.6  &    4902 &     3.20 &  $-$0.47 &  $-$0.25 &  $+$0.02 &  $-$0.23 &  $+$7.74 &  $+$0.13 \\
J000212.23$-$224138.9  &    5209 &     2.30 &  $-$2.52 &  $+$0.68 &  $+$0.01 &  $+$0.69 &  $+$6.61 &  $-$0.05 \\
J000216.68$-$245349.5  &    5018 &     1.75 &  $-$2.32 &  $-$0.06 &  $+$0.29 &  $+$0.23 &  $+$6.34 &  $+$0.42 \\
J000312.65$-$001504.4  &    4461 &     3.36 &  \nodata &  \nodata &  \nodata &  \nodata &  \nodata &  \nodata \\